\documentclass[12pt]{article}
\usepackage{amsmath, amsthm}
\usepackage{amsfonts}
\usepackage{indentfirst}
\textwidth 15cm \textheight 19cm \theoremstyle{theorem}
\newtheorem{theorem}{Theorem}

\newtheorem{propos}{Proposition}
\newcommand{\ds}{\displaystyle}
\begin{document}

\title{On the stability bounds in a problem of convection with uniform internal heat source}
\author{{\bf F.I. Dragomirescu$^{*}$, A. Georgescu$^{**}$}\\
\emph{{\small $^{*}$ Univ. "Politehnica"
of Timisoara, Dept. of Math., i.dragomirescu@gmail.com }}\\
\emph{{\small $^{**}$ Academy of Romanian Scientists,
adelinageorgescu@yahoo.com}}}

\date{}
\maketitle
\begin{abstract}
Two Galerkin methods are applied to a problem of convection with
uniform internal heat source are given. With each method
analytical results are obtained and discussed. They concern the
parameter representing the heating rate. Numerical results are
also given and they agree well with the existing ones.
\end{abstract}
\section{The physical problem}
\par  Natural convection
induced by an internal heat source is a phenomenon which has been
intensively studied, especially in order to point out its
influence on other processes. The motion in the atmosphere or
mantle convection are two among such phenomena \cite{Ve}. They
bifurcate from the conduction state as a result of its loss of
stability. A major importance is given to thermal convection
processes in terrestrial bodies driven by internal heat sources in
which the heat source is a function of time and, moreover, can
vary from one terrestrial body to another. In spite of their
importance, due to the occurrence of variable coefficients in the
nonlinear partial differential equations governing the evolution
of the perturbations around the basic equilibrium, so far these
phenomena were treated mostly numerically and experimentally. In
\cite{Drag1} we carried out a linear study for the eigenvalue
problem associated with the equations for a convection problem
with an uniform internal heat source in a horizontal fluid layer
bounded by two rigid walls \cite{Ve}. Our method was based on
Fourier series expansions for the unknown functions. Numerical
results and graphs were given showing a destabilizing effect of
the presence of the heat source. In \cite{Drag2} another two
methods based on Fourier series expansions (a Chandrasekhar
functions - based method and a shifted Legendre polynomials -
based method) were used to study analytically the  eigenvalue
problem deduced in \cite{Drag1}.
\par In \cite{TT} a linear stability analysis for a natural convection problem induced by internal heating is performed in
order to point out the effects of the heat distribution. This is a
function of both the critical Rayleigh number and the critical
wavenumber. Some non-uniform distributions were considered along
with the uniform one. It was shown that a concentration of the
heat source near the bottom boundary implies a decreasing of the
stability domain; namely it lowers the temperature difference at
which the convection sets in. The variation of the critical
wavenumber is small and there is only a slight influence of this
distribution on the size of the convection cells. When the heat
source is placed near the top boundary an enlargement of the
domain of stability occurs.
\par Another analytical study for a problem of convection in a fluid
saturated porous layer heated internally and in the presence of a
linearly varying gravity field is presented in \cite{Her}. It was
proved that the principle of exchange of stabilities holds as long
as the gravity field and the integral of the heat source have the
same sign. Convection in a medium with internal heat source was
also analyzed in \cite{AC} by linear stability methods and
nonlinear stability (energy type) methods. Numerical bounds for
the critical value of the control parameter, the Rayleigh number,
were given and the continuous dependence of the solution of the
initial boundary value problem on the internal heat source was
proved.
\par In \cite{Ve} a horizontal layer of viscous
incompressible fluid with constant viscosity and thermal
conductivity coefficients is considered. The performed numerical
investigation concerned the vertical distribution of the total
fluxes and their individual components for small and moderate
supercritical Rayleigh number in the presence of a uniform heat
source. In this context, the heat and hydrostatic transfer
equations  are \cite{Ve}
\begin{equation}
\label{eq:heat} \eta=k\dfrac{\partial^{2}\theta_{B}}{\partial
z^{2}},
\end{equation}
\begin{equation}
\label{eq:hydro} \dfrac{dp_{B}}{dz}=-\rho_{B}g,
\end{equation}
where $\eta=const.$ is the heating rate, $\theta_{B}$, $p_{B}$ and
$\rho_{B}$ are the potential temperature, pressure and density in
the basic state. In the fluid, the temperature at all point varies
at the same rate as the boundary temperature, so the problem is
characterized by a constant potential temperature difference
between the lower and the upper boundaries
$\Delta\theta_{B}=\theta_{B_{0}}-\theta_{B_{1}}$. Taking into
account (\ref{eq:heat}) this leads to the following formula for
the potential temperature distribution \cite{Ve}
\begin{equation}
\label{eq:potential_t}
\theta_{B}=\theta_{B_{0}}-\dfrac{\Delta\theta_{B}}{h}\Big(z+\dfrac{h}{2}\Big)+\dfrac{\eta}{2k}\Big[z^{2}-
\Big(\dfrac{h^{2}}{2}\Big)^{2}\Big].
\end{equation}
\par  In nondimensional variables the governing system of equations is
\begin{equation}
\label{eq:ecuatie_nond} \left\{
\begin{array}{l}
\dfrac{d{\bf U}}{dt}= -\nabla p'+\Delta {\bf U}+Gr \theta' {\bf k},\\
\textrm{div} {\bf U}=0,\\
\dfrac{d\theta'}{dt}=(1-Nz){\bf U}{\bf k}+Pr^{-1}\Delta \theta',
\end{array}
\right.
\end{equation}
where ${\bf U}=(u,y,w)$ is the velocity, $\theta'$ and $p'$ are
the temperature
 and pressure deviations from the basic state [6], $Gr$ is the Grashof number, $Pr$ is the Prandtl number and
 $N$ is a dimensionless parameter characterizing the heating (cooling) rate of the layer.
 \par The boundaries are considered rigid and ideal heat conducting, so the boundary conditions
 read
 \begin{equation}
 \label{eq:b_conditions}
 {\bf U}=\theta'=0 \textrm{ at } z=-\dfrac{1}{2} \textrm{ and } z=\dfrac{1}{2}.
 \end{equation}
 In
\cite{Drag1} in order to deduce the eigenvalue problem we
considered the viscous incompressible fluid confined into a
rectangular box bounded by two rigid walls:
  $V:0\leq x \leq a_{1}$, $ 0\leq y\leq a_{2}$, $-\dfrac{1}{2}\leq z\leq \dfrac{1}{2}$.
  We assumed that any unknown function in (\ref{eq:ecuatie_nond}) is of the form from
  \cite{Ge1}
 $$
 f(x,y,z)=\overline{F}(z)exp\Big(i\Big(\ds{2\pi m'\frac{x}{a_{1}}+2\pi n'\frac{y}{a_{2}}}\Big)\Big),
$$
 $m=\dfrac{2\pi m'}{a_{1}},
 n=\dfrac{2\pi n'}{a_{2}}$, where $a_{1}=\dfrac{L}{H}$, $a_{2}=\dfrac{l}{H}$, $L$ and $l$ are the box sizes.
Here $m'\geq 1$ and $n'\geq 1$ are the number of cells in the $x$
and  the $y$ direction.
\par Another possibility is to assume disturbances periodic in
$x$ (period $2\pi/\alpha$) and $y$ (period $2\pi/\beta$), with a
growth rate $\sigma$, also of the form
$$
 f(x,y,z)=\overline{F}(z)exp\Big(\sigma t+i\alpha x+i\beta y\Big).
$$
In this case, a subsequent investigation will concern the
condition in which the principle of exchange of stabilities is
valid.
\par In this paper, we treat only the stationary case
and this implies that the principle of exchange of stabilities is
valid. We complete our analytical study from \cite{Drag1},
\cite{Drag2} with some remarks on the spectral methods used to
solve the eigenvalue problem governing the linear stability of the
basic state for the convection problem with uniform internal heat
source.\par
 The eigenvalue problem associated with the
 equations (\ref{eq:ecuatie_nond})-(\ref{eq:b_conditions}) in a
 horizontal fluid layer bounded by two rigid walls, governing the stability of the basic motion against normal mode perturbations, deduced by
 us in \cite{Drag1} has the form
\par \begin{equation}
\label{eq:eigen1} \left\{
\begin{array}{l}
(D^{2}-a^{2})^{2}W=\Theta,\\
(D^{2}-a^{2})\Theta=-a^2 R(1-Nx)W
\end{array}
\right.
\end{equation}
with the boundary conditions
\begin{equation}
\label{eq:bc_eigen1} W=DW=\Theta=0 \textrm{ at }
x=\pm\dfrac{1}{2}.
\end{equation}
Here the Rayleigh number $R=Gr\cdot Pr$ represents the eigenvalue,
while $W, \Theta$, the amplitudes of the perturbations for the
velocity and the temperature field respectively, form the
corresponding eigenvector is $(W,\Theta)$.
\section{On the convergence of the Galerkin method}
In this section,  we reveal some aspects of the convergence  of
the Galerkin method,  one of the most used method for converting a
differential operator boundary value problem to a discrete one.
 \par There are more than one analytical
possibilities to solve the system
(\ref{eq:eigen1})-(\ref{eq:bc_eigen1}). However, some remarks on
the convergence of the system are in order. First, let us perform
a translation of variables $z=x+\dfrac{1}{2}$, such that the
problem (\ref{eq:eigen1}) becomes
\begin{equation}
\label{eq:eigen2} \left\{
\begin{array}{l}
(D^{2}-a^{2})^{2}W-\Theta=0,\\
(D^{2}-a^{2})\Theta+a^{2}R (N_{1}-Nz)W=0,
\end{array}
\right.
\end{equation}
with $N_{1}=1+\dfrac{N}{2}$ and the boundary conditions
\begin{equation}
\label{eq:bc_eigen2} W=DW=\Theta=0 \textrm{ at }z=0 \textrm{ and
}1.
\end{equation}\par The equations from
(\ref{eq:eigen2}) can be considered as a particular case of a more
general eigenvalue problem with variable coefficients \cite{DiP1}
\begin{equation}
\label{eq:eig_gen} \left\{
\begin{array}{l}
(D^{2}-a^{2})^{2}W= f(z)\Theta,\\
(D^{2}-a^{2})\Theta=-a^{2}Rg(z)W,
\end{array}
\right.
\end{equation} on $0\leq z\leq 1$.
\par The mathematical problem reads:\emph{ for given $f(z)$ and $g(z)$
(in our case $f(z)=1$ and $g(z)=N_{1}-Nz$) determine the minimum
real positive $R$ over all real positive} $a$ \emph{for which
there exists a nonnul solution of the system
(\ref{eq:eig_gen})-(\ref{eq:bc_eigen2}).}
\par Following Kolomy \cite{Ko} the convergence of the Galerkin method can
be considered for the sixth-order equation
$(D^{2}-a^{2})^{3}W=-a^{2}R(N_{1}-Nz )W$ obtained by eliminating
$\Theta$ between the two equations from (\ref{eq:eig_gen}). The
following result holds: \begin{propos} The operator
$L=(D^{2}-a^{2})^{3}$ is not symmetric in the sense of an
$L^{2}(0,1)$ inner product on a space of functions satisfying
$W=D^{2}W=(D^{2}-a^{2})^{2}W=0$ at $z=0,1$. \end{propos} In order
to prove Proposition 1, consider the inner product $(LW,W^{*})$ in
$L^{2}(0,1)$ with $W,W^{*}$ functions from $\mathcal{D}L$,
$$\mathcal{D}L:=\{U\in L^{2}(0,1)|U=D^{2}U=(D^{2}-a^{2})^{2}U=0 \textrm{ at }z=0,1\}.$$
The operator $L$ is said to be symmetric if $(LW,
W^{*})=(W,LW^{*})$ for any $W, W^{*}\in \mathcal{D}L$.  In our
case, by direct integration by parts it can be proven that
$(LW,W^{*})=(W,LW^{*})$. However, $W^{*}$ is not a function from
$\mathcal{D}L$, namely $W^{*}$ satisfies boundary conditions of
the type
\begin{equation}
\label{eq:w*} W^{*}=D^{2}W^{*}=D(D^{2}-a^{2})W^{*}=0 \textrm{ at }
z=0,1,
\end{equation}
whence Proposition 1. In \cite{Drag1} the quoted sixth order
equation together with the boundary conditions (\ref{eq:w*}) was
investigated  using spectral methods based on trigonometric
Fourier series and  good numerical results were obtained.
\par Consider the eigenvalue problem (\ref{eq:eigen1})-(\ref{eq:bc_eigen1}). Rescalling
(\ref{eq:eigen1}) by the factor $\dfrac{1}{\lambda}$,
$\lambda=a^{2}R$ the eigenvalue problem can be written in the form
$Aw-\lambda Kw=0$, with
\begin{equation}
\label{eq:operators} A=\begin{pmatrix} (D^{2}-a^{2})^{2}& 0\\
0& (D^{2}-a^{2})
\end{pmatrix}, \ \ \ K=\begin{pmatrix}
0&1\\
Nx-1&0
\end{pmatrix}.
\end{equation}
Here $w\in \mathcal{D}_{A}$, with $\mathcal{D}(A)$ the definition
domain of the matricial differential operator $A$ given by
$$\mathcal{D}_{A}:=\Big\{w=(W,\Theta)\in \Big(L^{2}\Big(-\dfrac{1}{2}\Big), \dfrac{1}{2}\Big)^{2}| W=DW=\Theta=0 \textrm{ at } z=-\dfrac{1}{2},\dfrac{1}{2}\Big\}.$$
The following convergence result was proved.
\begin{theorem}\cite{DiP1} Let $\lambda$ be a parameter in the
equation
\begin{equation}
\label{eq:th} Aw-\lambda Kw=0,
\end{equation}
where $A$ and $K$ are linear operators, and the domain of $A$,
$D_{A}$, is a linear manifold that is dense in a Hilbert space $H$
with the inner product $(\cdot, \cdot)$. Let $D_{A}$ be contained
in the domain of $K$, and assume that the following conditions are
fulfilled:
\begin{itemize}
\item[a)]the operator $A$ is a positive-definite, selfadjoint
operator; that is $(Au,u)>0$ and $(Au,v)-(v, Au)=0$;
\item[b)]the  operator $A^{-1}K$  can be extended  to a
completely continuous operator on the Hilbert space $H_{n}$, where
$H_{n}$ is the completion of $D_{A}$ under the norm
$(Au,u)^{1/2}$.
\end{itemize}
Then the Galerkin method for calculating the eigenvalues of
(\ref{eq:th}) is a convergent process in $H_{n}$.
\end{theorem}

Using the definitions of the matricial differential operators, all
the conditions of the theorem are satisfied so, the Galerkin
method for computing the eigenvalues of (\ref{eq:eigen1})
converges in the norm of $H_{0}$, with $H_{0}$ the Hilbert space
obtained by completing $\mathcal{D}_{A}$ (which is a preHilbert
space) to a Hilbert space.
\par {\bf Remark.}  Similarly, the convergence can be proved in
the case of $L^{2}(0,1)$.
\section{Galerkin type spectral methods}
\par  The expansion functions used for the unknown fields
encountered in various  convection problems from hydrodynamic
stability theory must have a  basic property: they must be easy to
evaluate. Trigonometric and polynomial functions have this
property. A second requirement is the completeness of the sets of
expansion  functions. This assures that each function of the given
space can be written as a linear combination of functions from the
considered set (or, more likely, as a limit of such a linear
combination). The Chebyshev polynomials, the Legendre polynomials,
the Hermite functions, the sine and cosine functions, satisfy this
condition.

\par In the Galerkin approach used here the basis (trial) functions satisfy the boundary
conditions. In this case, following Rama Rao \cite{RR}, the
simplest choice seems to be to write $W$ and $\Theta$ as
\begin{equation}
\label{eq:f_exp} W=\sum\limits_{m=0}^{\infty}a_{m}h_{1m}(x), \ \
\Theta=\sum\limits_{m=0}^{\infty}b_{m}h_{2m}(x),
\end{equation}
where $h_{1m}(x)=(1-4x^{2})^{m+2}$, $h_{2m}(x)=(1-4x^{2})^{m+1}$.
With this choice, the unknown functions $W$ and $\Theta$ satisfy
the boundary conditions (\ref{eq:bc_eigen1}). Replacing these
expressions in (\ref{eq:eigen1}) and imposing the condition that
the obtained equations are orthogonal to $h_{1n}(x)$ and
$h_{2n}(x)$ respectively, $n\in \mathbb{N}$ we obtained an
algebraic system in the unknown coefficients $a_{m}$ and $b_{m}$.
The condition that these coefficients are nonnull gives us the
secular (dispersion) equation. However, an important remark is in
order: the physical parameter $N$ representing the heating
(cooling) rate is missing from this equation.
\par Let us mention that the method is used in Ramma Rao \cite{RR}
in a convective instability problem of a heat conducting
micropolar fluid layer situated between two rigid boundaries. In
order to investigate the critical values of the Rayleigh number at
which instability sets in the most rough approximation is taken,
with only one term for each expression from (\ref{eq:f_exp}), so
the approximate values of $R$ are also crude. Nevertheless, in our
case, for this approximation, in the classical case of B\'{e}nard
convection, corresponding to $N=0$, the critical value of the
Rayleigh number $R$ is $R=1705. 715$ for $a=3.17$, which is a very
good approximation compared to the well-known value from
Chandrasekhar \cite{Ch}. We can conclude that this approximation
works with good results only in the classical case. \par A
mathematical explanation for the absence of the parameter $N$
could be that the chosen set of expansion functions introduced an
extraparity in the problem, leading to the loss of one of the
physical parameter, in this case the heating (cooling) rate $N$.
\par In \cite{Drag2} we considered also a basis of some
hyperbolic  functions for the expansion of the unknown function
$W$, i.e. $W=\sum\limits_{n=1}^{\infty}W_{n}^{1}C_{n}(x)$. For
this choice the physical parameter $N$ was not present in the
dispersion equation. This is why, we assume that a more suitable
choice is to consider the general case
$$W(x)=\sum\limits_{n=1}^{\infty}W_{n}^{1}C_{n}(x)+W_{n}^{2}S_{n}(x)$$
where $C_{n}$ and $S_{n}$ are the Chandrasekhar sets of functions
defined in \cite{Ch}
\begin{equation}
\label{eq:1funct}
\begin{array}{l}
\{C_{n}\}_{n\in\mathbb{N}},\ \ \
C_{n}(z)=\dfrac{\cosh(\lambda_{n}z)}{\cosh(\lambda_{n}/2)}-
\dfrac{\cos(\lambda_{n}z)}{\cos (\lambda_{n}/2)},\\
\\
\{S_{n}\}_{n\in \mathbb{N}},\ \ \
S_{n}(z)=\dfrac{\sinh(\mu_{n}z)}{\sinh(\mu_{n}/2)}-\dfrac{\sin
(\mu_{n}z)}{\sin(\mu_{n}/2)},
\end{array}
\end{equation}
with $\lambda_{n}$, $\mu_{n}$ given in \cite{Ch} by explicit
values for $n=1,2,3,4$ and by a recurrence relation for $n> 4$.
\par From $(\ref{eq:eigen1})_{2}$ we obtain
the expression of the unknown function $\Theta$,
$$\Theta(x)=-a^{2}R\sum\limits_{i=1}^{2}\Theta_{i}(x)+A\cosh(ax)+B\sinh(ax)$$
with $A,B$ deduced from the boundary conditions $\Theta\Big(\pm
\dfrac{1}{2} \Big)=0$. The functions $\Theta_{i}(x)$,
$i=1,2,...,4$, depending on the  coefficients $W_{n}^{1}$ and
$W_{n}^{2}$, have the form
$$
\left\{
\begin{array}{l}
\Theta_{1}(x)=\sum\limits_{n=1}^{\infty}\Big\{\dfrac{W_{n}^{1}(Nx-1)\cosh(\lambda_{n}x)}{(\lambda_{n}^{2}-a^{2})\cosh(\lambda_{n}/2)}-
\dfrac{2N\lambda_{n}W_{n}^{1}\sinh(\lambda_{n}x)}{(\lambda_{n}^{2}-a^{2})^{2}\cosh(\lambda_{n}/2)}\Big\};\\
\\
\Theta_{2}(x)=\sum\limits_{n=1}^{\infty}\Big\{\dfrac{W_{n}^{1}(Nx-1)\cos(\lambda_{n}x)}{(\lambda_{n}^{2}+a^{2})\cos(\lambda_{n}/2)}-
\dfrac{2N\lambda_{n}W_{n}^{1}\sin(\lambda_{n}x)}{(\lambda_{n}^{2}+a^{2})^{2}\cos(\lambda_{n}/2)}\Big\};\\
\\
\Theta_{3}(x)=\sum\limits_{n=1}^{\infty}\Big\{\dfrac{W_{n}^{2}(Nx-1)\sinh(\mu_{n}x)}{(\mu_{n}^{2}-a^{2})\sinh(\mu_{n}/2)}-
\dfrac{2N\mu_{n}W_{n}^{2}\cosh(\mu_{n}x)}{(\mu_{n}^{2}-a^{2})^{2}\sinh(\mu_{n}/2)}\Big\};\\
\\
\Theta_{4}(x)=\sum\limits_{n=1}^{\infty}\Big\{\dfrac{W_{n}^{2}(Nx-1)\sin(\mu_{n}x)}{(\mu_{n}^{2}+a^{2})\sin(\mu_{n}/2)}+
\dfrac{2N\mu_{n}W_{n}^{2}\cos(\mu_{n}x)}{(\mu_{n}^{2}+a^{2})^{2}\sin(\mu_{n}/2)}\Big\}.\\
\end{array}
\right.
$$
 \par
Let us replace this expression in $(\ref{eq:eigen1})_{1}$. The
orthogonality relation on $C_{m}$, $S_{m}$, $m\in \mathbb{N}$
imposed by the Galerkin procedure led us to an algebraic system
for the unknown coefficients $W_{n}^{1}$ and $W_{n}^{2}$
\begin{equation}
\left\{
\begin{array}{l}
\sum\limits_{n=1}^{\infty}W^{1}_{n}[(\lambda_{n}^{4}+a^{4})\delta_{nm}-2a^{2}T_{nm}]-2a^{2}W_{n}^{2}U_{nm}=\sum\limits_{i=1}^{4}C_{\Theta_{i}}+\sum\limits_{k=1}^{2}C_{m}^{k};\\
\sum\limits_{n=1}^{\infty}-2a^{2}V_{nm}W_{n}^{1}+W_{n}^{2}[(\mu_{n}^{4}+a^{4})\delta_{nm}]-2a^{2}P_{nm}=\sum\limits_{i=1}^{4}S_{\Theta_{i}}+\sum\limits_{k=1}^{2}S_{n}^{k},
\end{array}
\right.
\end{equation}
with
$$T_{nm}=(C_{n}'', C_{m}); \ U_{nm}=(S_{n}'',C_{m}); \ V_{nm}=(C_{n}'',S_{m}); \ P_{nm}=(S_{n}'', S_{m})$$
and
$$
\begin{array}{l}C_{m}^{1}=\ds\int_{-1/2}^{1/2}\cosh(ax)C_{m}(x);\
C_{m}^{2}=\ds\int_{-1/2}^{1/2}\sinh(ax)C_{m}(x);\\
S_{m}^{1}=\ds\int_{-1/2}^{1/2}\cosh(ax)S_{m}(x);\
S_{m}^{2}=\ds\int_{-1/2}^{1/2}\sinh(ax)S_{m}(x);\\
C_{\Theta_{i}}=\ds\int_{-1/2}^{1/2}\Theta_{i}(x)C_{m}(x);\
S_{\Theta_{i}}=\ds\int_{-1/2}^{1/2}\Theta_{i}(x)S_{m}(x).
\end{array}
$$
This time, the secular equation depends on $N$ and it follows from
the condition that not all these coefficients vanish. Numerical
values of the Rayleigh number are then obtained and displayed in
Table 1 in comparison with previous results.

\begin{center}
\begin{tabular}{|c|c|c|c|}
\hline $N$&$a^{2}$&$R_{a}- \cite{Drag1}$&$R_{a} - here$\\
\hline $0$&$9.711$&$1715.079324$&$1708.54$\\
\hline $1$&$9.711$&$1711.742588$&$1651.04$\\
\hline $2$&$9.711$&$1701.891001$&$1609.12$\\
\hline $1$&$10.0$&$1712.257687$&$1651.1$\\
\hline $4$ &$10.0$&$1664.341789$&$1560.8$\\
\hline $4$&$12.0$&$1685.422373$&$1739.2$\\
\hline $10$&$9.0$&$1482.527042$&$1366.02$\\
\hline $11$&$9.0$&$1446.915467$&$1366.05$\\
\hline $12$&$9.00$&$1411.401914$&$1354.7$\\
\hline
\end{tabular}\end{center}
\begin{center}
{\small {\bf Table 1. } Numerical evaluations of the Rayleigh
number for various values of the parameters $N$ and $a$.}
\end{center}
\par For the eigenvalue problem (\ref{eq:eigen2})-(\ref{eq:bc_eigen2}), in
\cite{Drag2}, in order to avoid the loss of the parameter $N$
different sets of orthogonal functions based on polynomials,
namely on shifted Legendre polynomials (SLP) on $[0,1]$ were
proposed. The method is similar to the one presented here. Instead
of $\{h_{1m}(x)\}_{m}$ and $\{h_{2m}(x)\}_{m}$ from
$L^{2}\Big(-\dfrac{1}{2}, \dfrac{1}{2}\Big)$, we used the
orthogonal sets from $L^{2}(0,1)$,
$$\{\beta_{m}(z)\}_{m}: \ \beta_{m}(z)=\ds\int_{0}^{z}\ds\int_{0}^{s}Q_{m+1}(t)dt
ds=\dfrac{1}{4}\Big[\dfrac{Q_{m+3}-Q_{m+1}}{(2m+3)(2m+5)}-\dfrac{Q_{m+1}-Q_{m-1}}{(2m+1)(2m+3)}\Big],$$
and $$\{\phi_{m}(z)\}_{m}: \
\phi_{m}(z)=\ds\int_{0}^{z}Q_{m}(t)dt=\dfrac{Q_{m+1}-Q_{m-1}}{2(2m+1)},$$
respectively,  with $Q_{m}$ the classical Legendre polynomials
defined on $[-1,1]$.
\par In this case, the expression of the secular equation contains
the physical parameter $N$, so good numerical evaluations of the
Rayleigh number for various values of $N$ and $a$ were obtained.
\par In \cite{DzR}, a general Galerkin type method is proposed for
the problem written in the general form
(\ref{eq:eig_gen})-(\ref{eq:bc_eigen1}). The unknown function
$\Theta$ is written as a Fourier series \cite{DzR} of the form
\begin{equation}
\Theta=\sum\limits_{m=1}^{\infty}A_{m}\cos(p_{m}z)+B_{m}\sin(q_{m}z),
\end{equation}
where $p_{m}=(2m-1)\pi$, $q_{m}=2m\pi$ which implies that $\Theta$
satisfies the boundary conditions (\ref{eq:bc_eigen2}). The
expression of $\Theta$, introduced in $(\ref{eq:eig_gen})_{1}$
leads to an expression of $W$ in the form $
W=\sum\limits_{m=1}^{\infty}A_{m}f_{m}(z)+B_{m}g_{m}(z) $ in which
the boundary conditions (\ref{eq:bc_eigen1}) are also considered
in order to find $A_{m}$ and $B_{m}$. However, in our case, the
function $f(z)$ is a constant one and the application of the
method in this form to (\ref{eq:eigen1})-(\ref{eq:bc_eigen1}) does
not lead to a correct expression of $W$.
\section{Conclusion}
In this paper a problem of convection with uniform internal heat
source is investigated. We complete a previous analytical study
\cite{Drag1}, \cite{Drag2} with some comments on  the choice of
the expansion functions and their importance for the convergence
of the Galerkin method. The importance of the form of the system
of ordinary differential equations which describe the eigenvalue
problem governing to the linear stability of the stationary
solution with respect to this choice is pointed out. We present
numerical results for the new introduced methods which are similar
to the ones obtained before.
\par The main conclusion of our analytical and numerical study performed in this paper and the previous ones is that
the choice of subspaces of trial functions with respect to whom
the approximation problems are solved influences the form of the
algebraic system and also the numerical evaluations. The good
numerical results obtained for small values of the spectral
parameter are justified by the accuracy of spectral methods.

\end{document}